\documentclass[longauth]{aa} 

\usepackage{graphicx}
\usepackage{txfonts}
\usepackage{amsmath}
\usepackage{xcolor}
\usepackage{ulem}

\usepackage[]{appendix}

\newcommand{\sntot}[0]{(\mathrm{S/N})_{\mathrm{tot}}}%
\newcommand{\snks}[0]{(\mathrm{S/N})_{\mathrm{KS}}}%

\usepackage{natbib,twoopt}
\usepackage[breaklinks=true]{hyperref} 
\bibpunct{(}{)}{;}{a}{}{,} 
\makeatletter
\newcommandtwoopt{\citeads}[3][][]{\href{http://adsabs.harvard.edu/abs/#3}%
{\def\hyper@linkstart##1##2{}%
\let\hyper@linkend\@empty\citealp[#1][#2]{#3}}}
\newcommandtwoopt{\citepads}[3][][]{\href{http://adsabs.harvard.edu/abs/#3}%
{\def\hyper@linkstart##1##2{}%
\let\hyper@linkend\@empty\citep[#1][#2]{#3}}}
\newcommandtwoopt{\citetads}[3][][]{\href{http://adsabs.harvard.edu/abs/#3}%
{\def\hyper@linkstart##1##2{}%
\let\hyper@linkend\@empty\citet[#1][#2]{#3}}}
\newcommandtwoopt{\citeyearads}[3][][]%
{\href{http://adsabs.harvard.edu/abs/#3}
{\def\hyper@linkstart##1##2{}%
\let\hyper@linkend\@empty\citeyear[#1][#2]{#3}}}
\makeatother

\begin{document} 

\title{K-Stacker, an algorithm to hack the orbital parameters of planets hidden in high-contrast imaging \thanks{Based on observations collected at the European Southern Observatory under programmes: 095.C-0298, 096.C-0241, 097.C-0865, 198.C-0209,
099.C-0127}} 

\subtitle{First applications to VLT/SPHERE multi-epoch observations}

\author{H. Le Coroller \inst{1}
          \and
          M. Nowak \inst{2, 3}
           \and 
           P. Delorme \inst{4}
           \and
          G. Chauvin \inst{4}
          \and 
          R. Gratton \inst{5}
            \and
          M. Devinat \inst{1}
          \and
           J. Bec-Canet \inst{1}
            \and
          A. Schneeberger  \inst{1} 
           \and
           D. Estevez \inst{6}
            \and
          L. Arnold \inst{7}
          \and
          H. Beust \inst{4} 
          \and
          M. Bonnefoy \inst{4}
          \and
          A. Boccaletti \inst{8}
           \and 
           C. Desgrange \inst{9, 10} 
           \and
           S. Desidera \inst{5}
           \and
           R. Galicher \inst{8}
           \and
          A.M. Lagrange \inst{4}
           \and
          M. Langlois \inst{11}
          \and 
          A.L. Maire \inst{12}
           \and
          F. Menard \inst{4}
          \and
          P. Vernazza \inst{1}
          \and
          A. Vigan \inst{1}
          \and
          A. Zurlo \inst{10, 13, 1}
           \and
             T. Fenouillet \inst{1}
          \and
          J.C. Lambert \inst{1}
          \and
          M. Bonavita \inst{5}
          \and
          A. Cheetham \inst{14}
           \and
          V. D'orazi \inst{5}
           \and
          M. Feldt \inst{15, 16}
          \and
          M. Janson \inst{15}
          \and
          R. Ligi \inst{17}
           \and
          D. Mesa \inst{5}
          \and 
          M. Meyer \inst{18}
          \and
          M. Samland \inst{15, 16}
          \and
          E. Sissa \inst{5}
          \and
          J.-L. Beuzit \inst{1}
			\and
			K. Dohlen \inst{1} 
			\and
			T. Fusco \inst{19}    
           \and
           D. Le Mignant \inst{1}
           \and
           D. Mouillet \inst{4}
           \and 
           J. Ramos \inst{15}
           \and
           S. Rochat \inst{4}
           \and
           J.F. Sauvage \inst{19} 
           }

   \institute{
   Aix Marseille Univ., CNRS, CNES, LAM, Marseille, France
    \email{herve.lecoroller@lam.fr} \\
    \and
    Institute of Astronomy, University of Cambridge, Madingley Road, Cambridge CB3 0HA, UK \\
    \and
    Kavli Institute for Cosmology, University of Cambridge, Madingley Road, Cambridge CB3 0HA, UK \\
      \and
      Univ. Grenoble-Alpes, CNRS, IPAG, F-38000 Grenoble, France \\
      \and
    INAF-Osservatorio Astronomico di Padova, Vicolo dell’Osservatorio 5, 35122 Padova, Italy \\
    \and
    LAPP UMR5814, 9 Chemin de Bellevue BP 110 Annecy-le-Vieux F-74941 ANNECY CEDEX, FRANCE \\
    \and 
    CFHT Corporation, 65-1238 Mamalahoa Hwy, Kamuela, Hawaii 96743, USA \\
   \and
    LESIA, Observatoire de Paris, CNRS, University Pierre et Marie Curie Paris 6 and University Denis Diderot Paris 7, 5 place Jules Janssen, 92195 Meudon, France \\
    \and
    Univ. Lyon, ENS de Lyon, Univ Lyon1, Lyon, France. \\
    \and
    N\'ucleo de Astronom\'ia, Facultad de Ingenier\'ia y Ciencias, Universidad Diego Portales, Av. Ejercito 441, Santiago, Chile\\
    \and
     CRAL, UMR 5574, CNRS, Université Lyon 1, 9 avenue Charles André, 69561 Saint Genis Laval Cedex, France; CRAL, UMR 5574, CNRS, Université Lyon 1, 9 avenue Charles André, 69561 Saint Genis Laval Cedex, France \\
    \and
    STAR Institute/Université de Liège, Allée du Six Août 19c, 4000, Liège, Belgium \\
    \and
    Facultad de Ingeniería y Ciencias,  Universidad Diego Portales, Av. Ejercito 441, Santiago, Chile \\
     \and
     Geneva Observatory, University of Geneva, Chemin des Maillettes 51, 1290, Versoix, Switzerland \\
    \and
     Max-Planck-Institut für Astronomie, Königstuhl 17, 69117, Heidelberg, Germany \\
     \and
     Department of Astronomy, Stockholm University, AlbaNova University Center, 10691, Stockholm, Sweden \\
     \and
      INAF-Osservatorio Astronomico di Brera, Via E. Bianchi 46, I-23807 Merate, Italy \\
     \and
     Department of Astronomy, University of Michigan, 1085 S. University Ave, Ann Arbor, MI, 48109-1107, USA\\
     \and
     DOTA, ONERA, Université Paris Saclay, F-91123, Palaiseau France \\    
             }

   \date{Received January 28, 2020; accepted April 21, 2020}

 
  \abstract
   { Recent high-contrast imaging surveys using the Spectro-Polarimetic High contrast imager for Exoplanets REsearch (SPHERE) or the Gemini Planet Imager (GPI), looking for planets in young, nearby systems showed evidence of a small number of giant planets at relatively large separation beyond typically 10 to 30 \,au where those surveys are the most sensitive. Access to smaller physical separations between 5 and 30 au is the next step for future planet imagers on 10 m telescopes and the next generation of extremely large telescopes in order to bridge the gap with indirect techniques such as radial velocity, transit, and soon, astrometry with \textit{Gaia}. In addition to new technologies and instruments, the development of innovative observing strategies combined with optimized data processing tools is participating to the improvement of detection capabilities at very close angular separation. In that context, we recently proposed a new algorithm, Keplerian-Stacker, combining multiple observations acquired at different epochs and taking into account the orbital motion of a potential planet present in the images to boost the ultimate detection limit. We showed that this algorithm is able to find planets in time series of simulated images of the SPHERE InfraRed Dual-band Imager and Spectrograph (IRDIS) 
   even when a planet remains undetected at one epoch.}
   { Our goal is to test and validate the K-Stacker algorithm performances on real SPHERE datasets, to demonstrate its resilience to instrumental speckles and the gain offered in terms of true detection. This will motivate future dedicated multi-epoch observation campaigns of well-chosen, young, nearby systems and very nearby stars to search for planets in emitted and reflected light, respectively, in order to open a new path concerning the observing strategy used with current and future planet imagers.  }
   {To test K-Stacker, we injected fake planets, and scanned the low S/N regime in a series of raw observations obtained by the SPHERE /IRDIS instrument in the course of the SHINE survey. We also considered the cases of two specific targets intensively monitored during this campaign: $\beta$ Pictoris and HD\,95086. For each target and epoch, the data were reduced using standard angular differential imaging processing techniques, then recombined with K-Stacker to recover the fake planetary signals. In addition, the known exoplanets $\beta$ Pictoris b and HD\,95086 b previously identified at lower S/N in single epochs have also been recovered by K-Stacker.}
  {We show that K-Stacker achieves high success rate of $ \approx 100 \%$ when the S/N of the planet in the stacked image reaches $\approx 9$. The improvement of the $S/N$ ratio goes as the square root of the total exposure time contained in the data being combined. At $\rm{S/N} < 6-7$, the number of false positives is high near the coronagraphic mask, but a chromatic study or astrophysical criteria can help to disentangle between a bright speckle and a true detection. During the blind test and the redetection of HD\,95086\,b, and $\beta$ Pic b, we highlight the ability of K-Stacker to find orbital solutions consistent with the ones derived by the state of the art MCMC orbital fitting techniques, confirming that in addition to the detection gain, K-Stacker offers the opportunity to characterize the most probable orbital solutions of the exoplanets recovered at low signal to noise.}
	{}

   \maketitle
%

\section{Introduction}

\qquad Most of the $4100$ exoplanets detected to date have been found using indirect methods, such as the radial velocity technique and photometric transits. It is indeed extremely difficult to detect the planet light that is drowned in the much brighter diffracted light from its host star. Jupiter and Earth like planets are about $10^8$ to $10^{10}$ fainter than their parent star in the visible band.\\
\qquad However, thanks to a combination of eXtreme Adaptive Optics (ExAO), innovative coronagraphs, differential imaging, and sophisticated post-processing algorithms, direct imaging instruments have been able to detect and characterize young giant planets at large separation ($> \sim$10 au).
 Across two decades of exoplanetary science in direct imaging, dozens of dedicated surveys have been
carried out around young, nearby stars \citepads{2018SPIE10703E..05C}. They led to the discovery of
the first planetary mass companions in the early 2000's, at large distances $\geq 100$ au. The implementation
of differential techniques, starting in 2005, enabled the breakthrough discoveries of closer and lighter planetary
mass companions like HR 8799 bcde (10, 10, 10 and $7$  M$_{\mathrm{Jup}}$
 at respectively 14, 24, 38 and 68 au; Marois et al. \citeyearads{2008Sci...322.1348M}, \citeyearads{2010Natur.468.1080M}), $\beta$ Pictoris b ($8$ M$_{\mathrm{Jup}}$ at 8 au; \citeads{2009A&A...506..927L}),
 Fomalhaut~b ($<1$~M$_{\rm{Jup}}$ at 177~au; \citeads{2008Sci...322.1345K}; still debated), HD\,95086\,b ($5$~M$_{\rm{Jup}}$ at 52~au; \citeads{2013ApJ...779L..26R}). 
 
The current generation of ExAO planet imagers 
(GPI, \citeads{2014PNAS..11112661M} ; SCExAO, \citeads{2015PASP..127..890J}; SPHERE, \citeads{2019A&A...631A.155B})
are now equipped with integral field spectrographs offering exquisite near-infrared spectra of young
giant planets to unveil the physical processes at play in their atmospheres and a link to their mechanisms of
formation. For the first time, these instruments have reached a contrast level of $\approx 10^{-5}$ at a separation of about $200$ to $900$\,mas, enabling the detection of the new young planets : 51\,Eri\,b ($2$~M$_{\rm{Jup}}$ at 13~au; \citeads{2015Sci...350...64M}), HIP\,65426\,b ($9$~M$_{\rm{Jup}}$ at 92~au; \citeads{2017A&A...605L...9C}), and PDS\,70\,b ($9$~M$_{\rm{Jup}}$ at 29~au; \citeads{2018A&A...617A..44K}). \\
An important finding from these high-contrast imaging surveys in the past years has been the low occurrence rate of giant planets beyond 30\,au (${0.6}_{-0.5}^{+0.7}\% $, see \citeads{2016PASP..128j2001B}). 
Today, the GPIES and SHINE large surveys of about 600 observed stars indicate that this scarcity extends down to 10\,au  (\citeads{Nielsen2019}; Vigan et al. 2020, submitted), suggesting that the bulk of the giant planet population is located between typically 1 and 10\,au. 

A prime goal of the future surveys will be also to bridge the gap with indirect techniques by imaging young Jupiters down to the snowline at about 3-5\,au, depending on stellar type. The next generation of instruments like SPHERE+ \citepads{2020arXiv200305714B} 
aim at reaching contrasts of at least $10^{-5}$ at $\approx 100\, mas$, which represents an improvement of a factor of 3 in terms of  angular separation with respect to SPHERE \citepads{2019A&A...631A.155B}. 
K-Stacker, together with the SPHERE+ capability, has the potential to achieve the core of the Jupiter-mass planets population at $\approx 3\, au$ (i.e. $\approx 10^{-6}$ at $\approx 60\, mas$) in $10-100$ hours of exposure time, by providing an additional factor of $3-10$ in contrast.
In terms of observing strategy, for planets at less than 10\,au from a star at $10$ to $20\,$pc, the orbital motion becomes comparable to the width of a 10 m telescope diffraction-limited point spread function (PSF) in about 30 days. 
 Taking into account observing constraints and weather statistics (higher contrast can be reached only during the best nights with a seeing  $< 0.6''$), multiple observations of very interesting young, nearby systems are usually spread over several days, months and years 
 (see case of $\beta$ Pictoris, \citeads{2019A&A...621L...8L}). In this case, the orbital motion of the potential planets makes a simple co-addition of the different images sub-optimal in terms of pure detection, if not impossible.
 
The Keplerian motion of the planet has to be taken into account during the combination. On the Extremely Large Telescopes (ELTs), the situation will be even worse. The PSF will be $4$ to $5$ times smaller than with the 10 m telescope class (VLT, Keck, Gemini, Subaru, LBT, LCO, etc.), and these ELTs will be used to search for planets at very small separations (below 10\,au) already with the first-light instruments \citepads{2018arXiv181002031C}. In this case, the Keplerian motion could become non-negligeable in a matter of a few days only \citepads{2013ApJ...771...10M}.

Following similar principles previously applied to the search of new moons in solar systems like Hippocamp, the seventh innermost moon of Neptune \citepads{2019Natur.566..350S}, the Keplerian-Stacker (K-Stacker) algorithm described by \citetads{2015tyge.conf...59L} is an observing strategy and method of data reduction applied to nearby stars, and that consists in combining high contrast images recorded during different nights, accounting for the orbital motion of the putative planet that we are looking for. Even if an individual image does not reveal the planet, we showed that an optimization algorithm like K-Stacker can be used to properly align the images according to Keplerian motion (for instance, $10$ to $50$ images taken over the course of several months or years). The resulting gain in signal-to-noise ratio (S/N) can allow for the detection of planets otherwise unreachable. This method can be used in addition to the angular differential imaging (ADI) and spectral differential imaging (SDI) techniques \citepads{1999PASP..111..587R, 2006ApJ...641..556M} or any other high contrast data reduction methods to further improve the global detection limit. As a byproduct of the optimization algorithm, K-Stacker also directly provides the orbital parameters of the detected planets.\\
Ultimately, the main goal of K-Stacker would be to directly drive the observing strategy and scheduling of current and future planet imagers in which exposures would be split over several nights to maximize the detection performances. In Paper I \citepads{2018A&A...615A.144N}, by using simulated VLT-SPHERE observations, we have shown that when the total number $n$ of available images is large enough to get $\sqrt{n}\times{}$ \text{(S/N)} $\ge{} 7$ (where  \text{(S/N)} is the Signal to Noise levels in individual frames), the K-Stacker algorithm is able to detect the planet with a high level of reliability $> 90\%$. The number of false positives were low but the simulated images did not reproduce instrumental speckle noise and angular spectral differential imaging (ASDI) reductions. The main goal of this paper is to validate the K-Stacker algorithm on real data obtained on sky reduced by the most recent algorithms (PCA ADI and ASDI).

In Section \ref{instrumentation_obs_used}, we describe the observations used in this paper, which come from the SPHERE SHINE survey \citepads{Chauvin2017b}.
In Section \ref{real data blind test and search}, we present the results of a blind test where K-Stacker was used to search for fake planets hidden in real IRDIS observations reduced by a PCA ADI algorithm. 
We also study the capability of K-Stacker to recover the correct orbital parameters despite the typical errors encountered in real data, such as instrumental true north offsets, or stellar mass and distance uncertainties.
In Section \ref{search for shine companions}, we show that K-Stacker is able to recover the known companions $\beta$ Pictoris b and HD 95086 b, and show that the orbital parameters retrieved by K-Stacker are in agreement with the values found in the literature.
We present in Section \ref{search for new planets} the first K-Stacker searches for new planets around HD 95086 and $\beta$ Pictoris, two targets which have been repeatedly observed during the SHINE survey. 
In Section \ref{discussion}, a discussion on the strategy of future K-Stacker observations is done. Section~\ref{conclusion} gives our final conclusions.


\section{Description of the observations used in this paper} \label{instrumentation_obs_used}

All the observations used in this paper come from the SHINE survey done with the Spectro-Polarimetic High contrast imager for Exoplanets REsearch (SPHERE) instrument at the focal plane of the VLT-UT3  \citepads{2019A&A...631A.155B}.
The SHINE program \citepads{2017A&A...605L...9C} is a very high-contrast near-infrared survey of more than 600 young, nearby stars aimed at searching for and characterizing new planetary systems. 
The goal of this project is also to find statistical constraints on the rate, mass and orbital distributions of the giant planet population at large orbits. 
Even if the SHINE observations have not been organized for the K-Stacker 'philosophy' where we plan to observe less stars but over more epochs, the number of observed targets reduced homogeneously by the SPHERE Data Center \citepads{2017sf2a.conf..347D, 2018AA...615A..92G} 
allow to test K-Stacker for the first time in real conditions.

SPHERE includes an extreme adaptive optics system, several types of coronagraphs, and three sub-systems (IRDIS, IFS and Zimpol). 
In this paper we use observations coming from the Integral Field Spectrograph (IFS) and the Infra-Red Dual-band Imager and Spectrograph (IRDIS), that were designed to cover the near-infrared range for an efficient search of young planets \citepads{2019A&A...631A.155B}. In  the  SHINE  survey,  two  configurations  were  used : APO1-ALC2 (N-ALC-YJH-S) coronagraph with a focal plane mask of a diameter of 185 mas, and the APO1-ALC3  (N-ALC-Ks) coronagraph with a focal plane mask of a diameter of 240 mas respectively optimized for the IRDIFS and IRDIFS-EXT modes \citepads{2019A&A...631A.155B}. In these modes, IRDIS and IFS work simultaneously (\citeads{2008epsc.conf..875C}, \citeads{2014A&A...572A..85Z}) and IRDIS is used in a Dual Band Imaging configuration (\citeads{2008SPIE.7018E..59D}, \citeads{2010MNRAS.407...71V}). 

All the IRDIS observations used in the blind test described in Section \ref{real data blind test and search} were acquired using the APO1-ALC2 coronagraph.
In Section \ref{search for shine companions} and \ref{search for new planets}, we also use the observations on two emblematic stars HD95086 and $\beta$ Pictoris that have been observed regularly in the SHINE survey using the APO1-ALC2 and ALC3 coronagraph (see Table \ref{tab:observations}) in order to constrain the orbital parameters
 of the known planets b around these targets.

\section{K-Stacker computation on real SHINE data} \label{real data blind test and search}
%
 \subsection{Set-up of the blind test} \label{blind test}
 
  In order to extend the demonstration of the K-Stacker algorithm on simulated IRDIS datasets presented in Paper I, we performed a new analysis on real IRDIS observations obtained during the SHINE survey. 
  
  The methodology followed during this new blind experiment is similar to the one discussed in Paper I, with the exception of an additional PCA ADI reduction step. To stay as close as possible to the conditions of Paper I, and to demonstrate the true potential of K-Stacker, we injected planets in fake "K-Stacker runs", each made of 10 observations taken from the SHINE survey. As most stars have not been observed that many times during the survey, we created the K-Stacker runs by combining observations acquired on different but similar targets. In particular, we were careful in selecting observations of stars with similar magnitudes, and acquired in similar conditions (seeing, AO performances, etc.). We create a total of 5 such fake K-Stacker runs, using a total of 50 different images from the survey. 
 
 \begin{table}[!h]
\begin{center}  
\caption{Parameters used to inject the planet in the 30 K-Stacker runs of 10 observations of our blind experiment. We chose the same ranges than in paper I.}
  \begin{tabular}{c l l}
    \hline
    \hline
    Parameter & Range & Distribution\\
    \hline
    $M_{\text{star}}$ & $1$ M$_\odot$ & fixed value \\
    $d_{\text{star}}$ & 10 pc & fixed value \\    
    $a$ & [2 au, 7.5 au] & uniform \\
    $e$ & [0, 0.5]  & uniform \\
    $t_0$ & [-20 yr, 0 yr] & uniform \\
    $\Omega$ & [-180 deg, 180 deg] & uniform \\
    $i$ & [0, 180 deg] & uniform \\
    $\theta_0$ & [-180 deg, 180 deg] & uniform \\
    \hline
  \end{tabular}
  \label{tab:parameters}
\end{center}  
\end{table}

 Fake planets are then injected in the raw observations of each run by doing the following:
 \begin{enumerate}
 \item{Randomly draw one run in which no planet is injected.}
 \item{For each of the 4 other runs, draw a set of random orbital parameters (see Table~\ref{tab:parameters} for an overview of orbital laws used), and inject the planet in each raw observation of the run according to the orbit drawn. The target star is assumed to have a mass of $1$ M$_\odot{}$, and to be located at 10 pc. The planet is injected at a random contrast uniformly drawn in the range $[5\times{}10^{-6}, 4\times{}10^{-7}]$.}
 \item{Reduce each observation of each run using the PCA ADI tools of the SPHERE Data Center \citepads{2017sf2a.conf..347D, 2018AA...615A..92G}.}
 \end{enumerate}
 
The process is repeated 6 times, in order to create a total of 30 fake K-Stacker runs, in which 6 have no planets, and 24 have a planet injected randomly. The resulting runs were then classified based on the perfectly recombined $\sntot$ ratio of the injected fake planets. Note that, the S/N definition when using K-Stacker can be confusing, as we need to distinguish 3 different values: the S/N ratio in each individual ADI observation of the run (simply S/N), the S/N after the K-Stacker recombination $\snks$, and the optimal S/N that could be achieved if the orbit was perfectly known, and the images perfectly recombined $\sntot$. The 30 runs were then divided into the three following Groups:
 
 \begin{itemize}
 \item{Group I, composed of 13 runs for which $\sntot < 2$ for the injected fake planets (or non injected fake planets). These planets can be considered as undetectable i.e. equivalent to no planet injected.}
 \item{Group II, composed of 9 runs for which $2<\sntot < 12$ for the injected fake planets (low-S/N regime)}
 \item{Group III, composed of 8 runs for which $\sntot> 12$ for the injected fake planets and therefore easily detectable.}
 \end{itemize}

\subsection{Blind check} 
\label{result_blind_test}

\begin{figure}[h]
   \includegraphics[width=10cm]{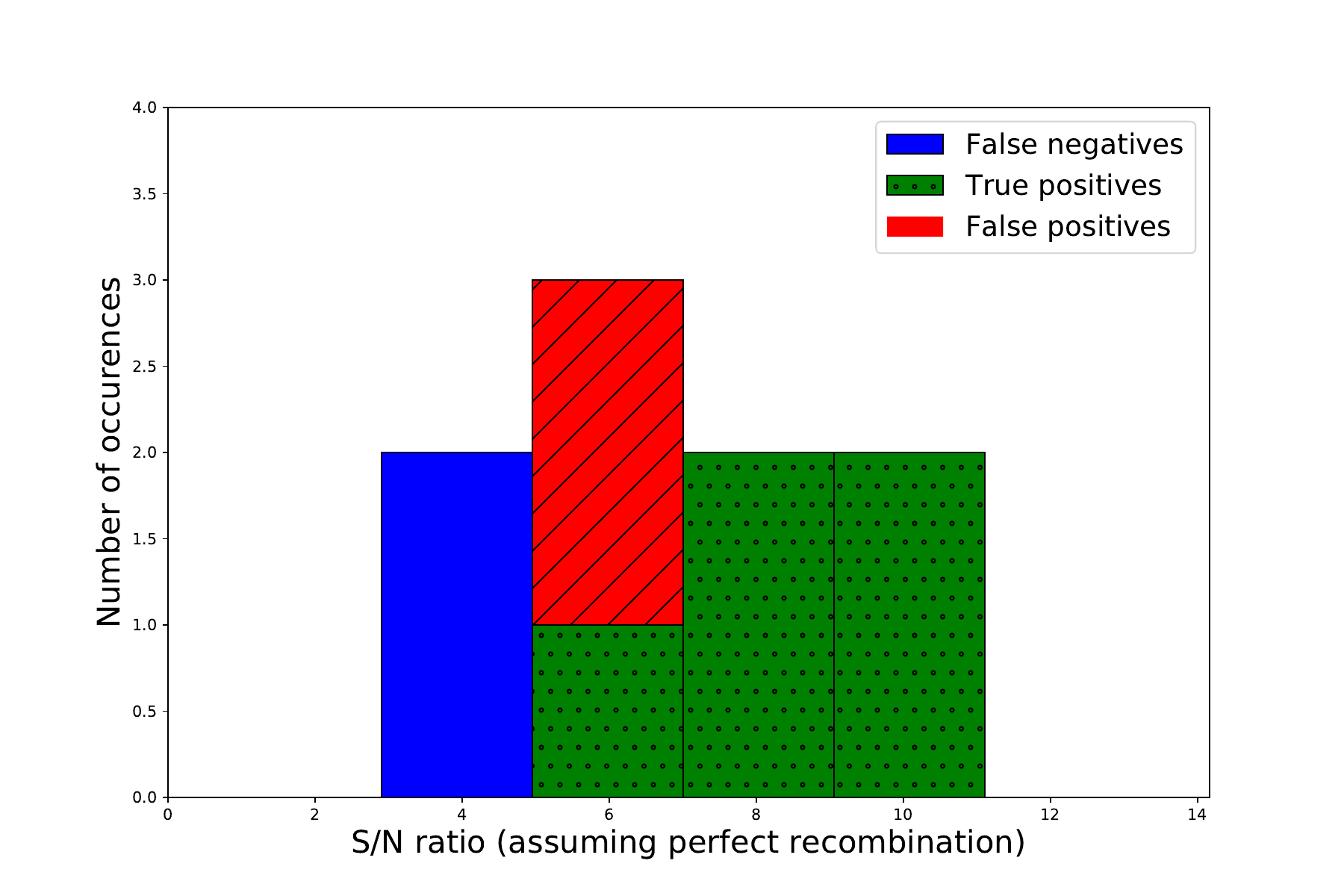}
   \caption{Distribution of the planet candidates found and missed as a function of the $\sntot$ given a perfect recombination of the images. The true negatives, all grouped at S/N=0 are not shown.}
     \label{Fig2}
\end{figure} 

\begin{figure}[!h]
   \includegraphics[width=8cm]{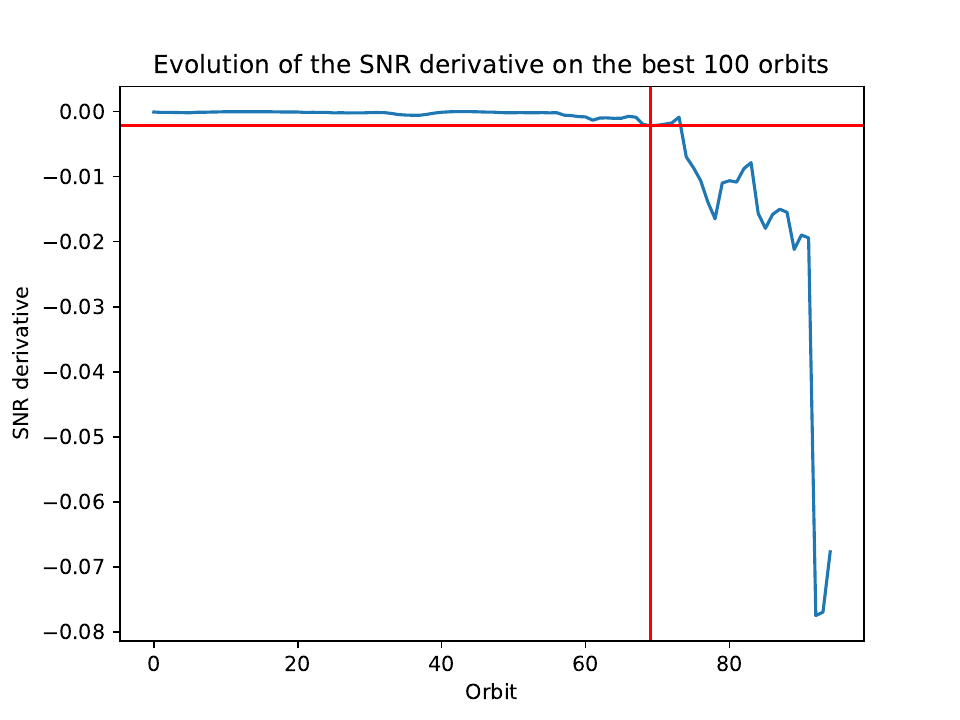}
   \caption{K-Stacker $\snks$ derivative in function of the orbit number sorted by signal to noise.}
     \label{Fig3}
\end{figure} 

Following the principles of Paper I, after running the K-Stacker algorithm, we asked to an independent observer (not aware of the presence or not of fake planets in the images, of the injection phase, the S/N or the orbital parameters) to check the final K-stacker recombined solutions and to assign for each run one of the three flags: "no detection", "planet candidate", and "possible candidate, more observations required".

Among the thirteen sets of Group I (i.e. $\sntot<2$), eight have been correctly flagged by the observer as no detection i.e. true negatives; The five other solutions were flagged as "possible candidates, more observations needed". Due to the nature of the blind test performed, in which the same limited number of high-contrast observations were used multiple times, all the runs are not fully independent, and these 5 cases actually correspond to two truly independent candidates emerging from the noise. The typical K-Stacker S/N ratio for these five cases was $\snks \approx 6$, and although the observer did not claim a detection from these runs, in reality, they would probably have led to more observation time being spent on the targets. We consider these two cases as false positives.

\begin{figure}[h]
   \includegraphics[width=9cm]{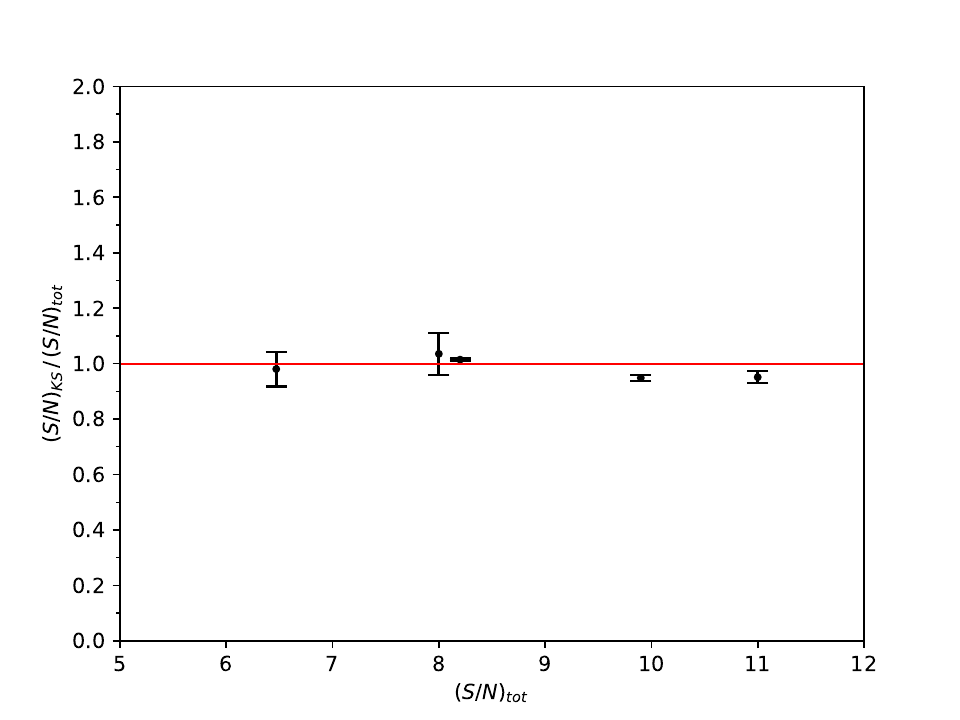}
   \caption{K-Stacker signal to noise $\snks$ divided by $\sntot$ of a perfect recombination, in function of $\sntot$.}
     \label{Fig:snr_ks_tot}
\end{figure} 

\begin{figure*}[!h]
   \includegraphics[width=18cm]{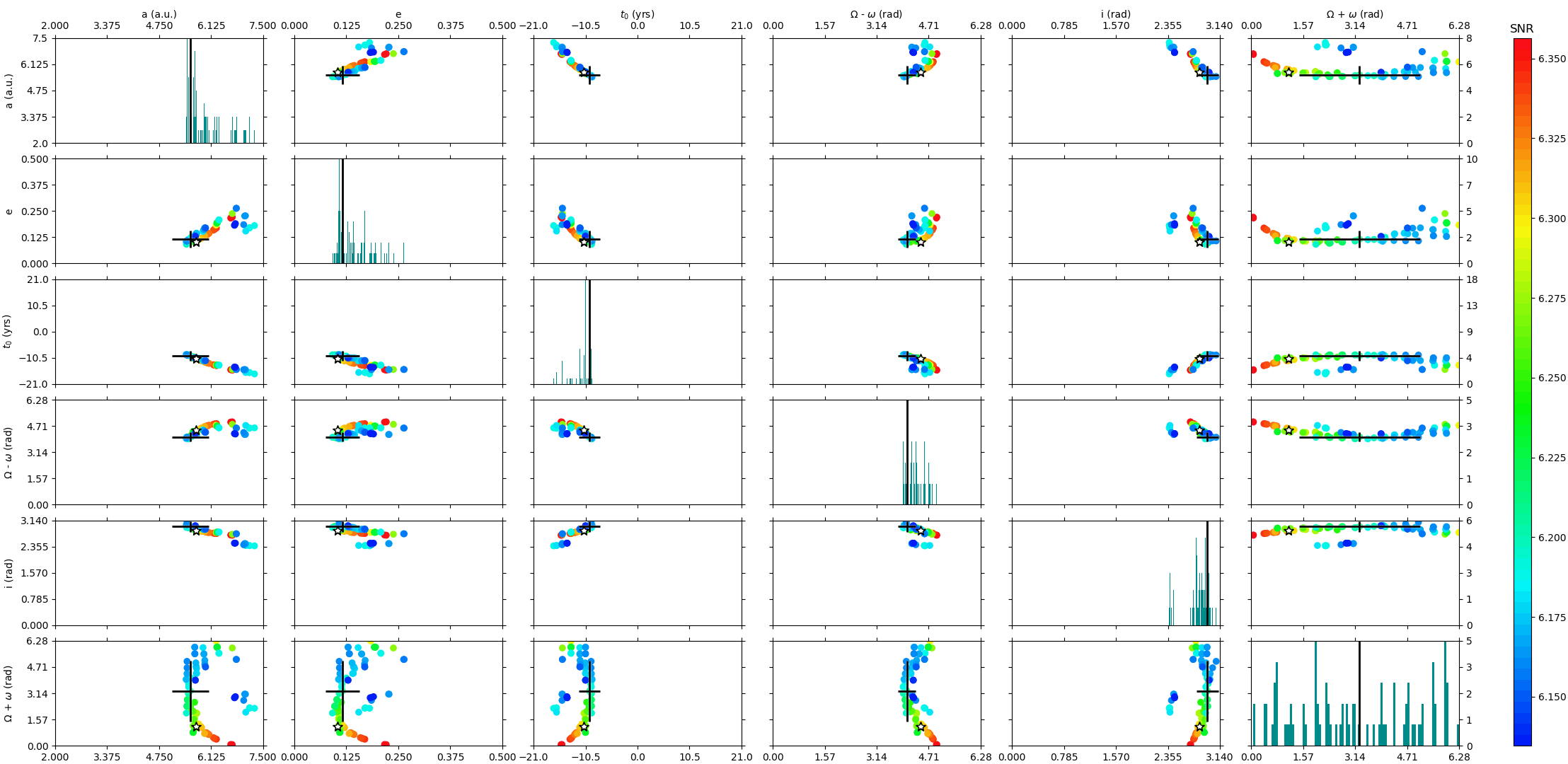}
   \caption{Example of orbital parameters coming from one true positive solution of the SHINE blind test (see Sect. \ref{result_blind_test}). In each sub-plot, each point corresponds to the parameters of one orbit found by  K-Stacker with the color of its $\snks$ value. In the diagonal, we show the histograms for each orbital parameter. 
   The dark cross corresponds to the mean of the K-Stacker orbits. The size of the dark cross is one sigma of each orbital parameter. The star is at the position of the orbital parameters used to inject the fake planet.} 
     \label{Fig_orbit_dc}
\end{figure*}

Among the nine runs of Group II (i.e. $2<\sntot<12$), five solutions were flagged as true detections by the observer and were indeed true positives. Two solutions at $\sntot=3.3$ and $\sntot=3.6$ have been found at $\snks \approx 6$, and flagged by the observer as possible detections, with more observations required. These two solutions correspond to the same false positives than described above. Two other solutions at $\sntot=3$ and $\sntot=4$ have been flagged as no detection by the observer, although the K-Stacker $\snks$ was found to be between 7 and 8. In these two last cases the orbits found by K-Stacker pass well by the true positions of the fake planet only in 3 images (epochs)  and not along the other last epochs. The contribution from the fake planet to the $\snks$ explains the relatively high values obtained. We count these ambiguous cases as false negatives. Finally, all the eight runs of Group III (i.e. $\sntot > 12$) have been flagged by the observer as detections and are indeed true positives.

In Fig.\ref{Fig2}, we report the histogram of the planets found and missed from Groups I and II. For clarity, solutions of Groups III with $\sntot > 12$ are not shown. All these solutions with $\sntot > 12$ correspond to true positives. Even if the statistics are poorer than in Paper I, we still find that K-Stacker is able to recover all the planets with $\sntot>6.5$ i.e. S/N $\simeq{}2$ in individual observations. With only five truly independent runs of 10 observations used to create the 30 fake K-Stacker runs, it is difficult to draw solid conclusions regarding the false positive rate. One of the two false alarms of Fig. \ref{Fig2} was found to be very close to the coronagraphic mask, where the false positive probability may be significantly higher (see also Section \ref{search for new planets}). However, in every cases, the observer did not claim a detection from the available observations, but merely suggested that the K-Stacker solution was possibly a planet, and requested more observations.

\subsection{Extracted orbital parameters} \label{orbit}

The result of the K-Stacker algorithm is a list of orbits sorted by signal to noise (see Paper I for detail). The signal to noise of the $\approx 50 - 100$ first orbits is a relatively flat function before decreasing by steps (see example in Fig.\,\ref{Fig3}). For our study of the orbital parameters, we keep only the orbits before this drop in $\snks$ i.e. 68 first orbits in the example of Fig. \,\ref{Fig3}. We have checked that for the true positives of the SHINE blind test the first K-Stacker solutions before the decrease of $\snks$ always contained a set of orbital parameters that pass well by the orbit of injection (see Fig \,\ref{Fig_orbit_dc}). We show the orbital parameters in 2D maps (Fig.\ref{Fig_orbit_dc} ) to be able to compare the K-Stacker solutions with the MCMC technique on the positions \citep{2016A&A...587A..89B}.

We also give a 'mean' solution with its standard deviation (black cross in Fig. \ref{Fig_orbit_dc}).
For all the planets detected by K-Stacker in the SHINE blind test (true positives in green of Fig. \ref{Fig2}), the mean solution of the orbital parameters is at maximum 3 standard deviations from the parameters of injection. 
This mean orbit always passes at less than $\approx 0.6$ pixels from the true positions of the fake planet in the images (see Tab. \ref{Tab_orbits}). 
Fig. \ref{Fig:snr_ks_tot} shows also that the K-Stacker signal to noise $\snks$ found with the mean solution of the orbital parameters is equal within the error bars to the $\sntot$ of a perfect recombination.
To conclude, K-Stacker has well recovered the orbital parameters although the planet has travelled over a maximum of $15-38\%$ of its total orbital period (see Fig.\,\ref{Fig_orbit_dc} and Table\ref{Tab_orbits}).

\subsection{K-Stacker tolerances on the errors of real data} \label{real data error}


To limit the computation time, K-Stacker does not include true north  error, or stellar mass and distance as free parameters. These values are fixed values set by the user. In this Section, we investigate the capability of K-Stacker to converge despite possible errors on these parameters. 

\subsubsection{Tolerance on the stellar mass error and impact on the orbital parameters}

To test the robustness of K-Stacker to errors on the stellar mass, we performed an experiment, in which we kept the mass of the target star to a fixed value $M = 1$ M$_\odot{}$ when calculating the orbit of the injected planets, but changed the mass used by K-Stacker to $M+\delta{}M$. 
For a circular face-on orbit of semi major-axis $a$, the orbital velocity is given by:

\begin{equation}
V_\mathrm{orb} = \sqrt{\frac{GM}{a}}
\end{equation}
\noindent{}Thus, a small error $\delta{}M$ on the mass $M$ of the star directly translates to an error $\delta{}V_\mathrm{orb}$ on the velocity given by:
\begin{equation}
\delta{}V_\mathrm{orb} = \frac{1}{2}\sqrt{\frac{G}{a}}\frac{\delta{}M}{{M}^{1/2}}
\end{equation}
\noindent{}This error on the orbital velocity will lead to a K-Stacker estimate of the position in each image which "drifts" with time, compared to the real position of the planet. Assuming that the algorithm can always play on appropriate parameters to minimize the total error by properly aligning the center of the time series, we can calculate the typical mean position error $\Delta{}X_\mathrm{mean}$ for a sequence of $N$ observations acquired at a regular time interval $\Delta{}T$ over a total period $T=(N-1)\Delta{}T$:
\begin{equation}
\Delta{}X_\mathrm{mean} = \frac{1}{N}\sum_{k=1}^{k=N}k\Delta{}T\delta{}V_\mathrm{orb} = \frac{T}{2}\delta{}V_\mathrm{orb}
\end{equation}
\noindent{}For a star at a distance $d$, the associated angular error is then:
\begin{equation}
\Delta{}\theta_\mathrm{mean} = \frac{1}{4}\frac{T}{d}\sqrt{\frac{G}{a}}\frac{\delta{}M}{{M}^{1/2}}
\label{eq:delta_mass}
\end{equation}

\begin{figure}[!h]
\begin{center}
   \includegraphics[width=9cm]{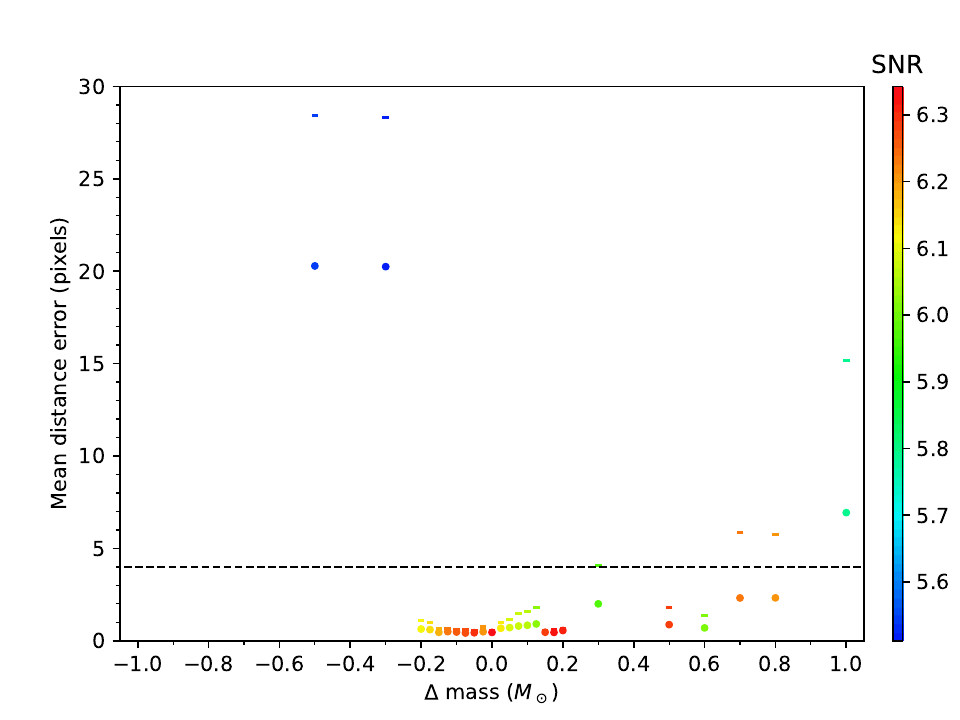}
   \caption{Results of the K-Stacker algorithm when including an error on the stellar mass used by the algorithm to calculate the orbital positions at each epoch. The dots (resp. dashes) show the mean (resp. maximum) distance between the true position and the position found by K-Stacker in all the images as a function of the error on the mass. These dots are color-coded to indicate the corresponding $\snks$. The black dashed line indicates the size of the FWHM of the instrumental PSF (i.e. K-Stacker has missed the planet at least in one image when a small line is above the dashed line).}
     \label{Fig:dist_delta_mass}
     \end{center}   
\end{figure}

\begin{figure}[!h]
\begin{center}
   \includegraphics[width=8cm, height=21cm]{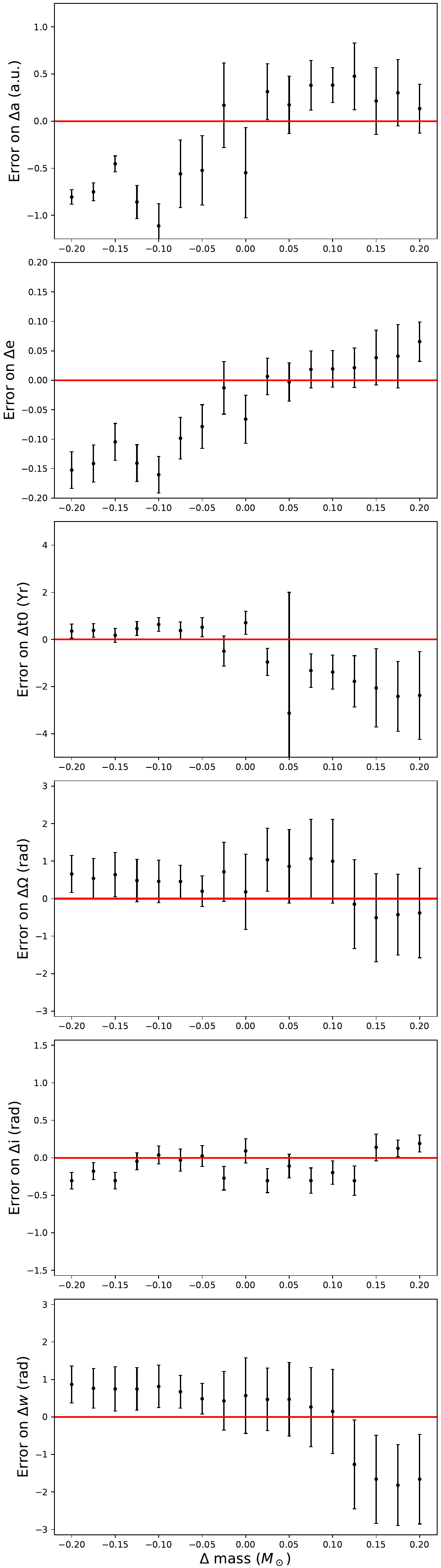}
   \caption{Difference between the orbital parameters found by K-Stacker and the real values of injection, as a function of the error on the stellar mass.}
     \label{Fig:orbit_param_delta_mass}
     \end{center}   
\end{figure}

With $d=10\,\mathrm{pc}$, $T=3\,\mathrm{yr}$, $a\simeq{}5$ au, and $M = 1\,$ M$_\odot{}$, an error of $0.1\,$ M$_\odot{}$ on the mass of the star leads to a mean angular error of 21~mas, similar to the size of the SPHERE PSF. For most of the stars of the SHINE survey the mass is known with an accuracy better than $10\, \%$  (\citeads{2015A&A...573A.126D}), so the effect should remain limited. 
Nevertheless, to study the impact of this error on the performance of K-Stacker, we used the algorithm on a fake run while explicitly adding an error on the stellar mass.
In Figure~\ref{Fig:dist_delta_mass}, we give the mean distance between the solution found by K-Stacker and the injected position of the planet, as a function of the error on the stellar mass $\delta{}M$. 
Interestingly, K-Stacker is able to tolerate relatively large error on the stellar mass ($-0.2<\delta\,M< 0.2$), which should in principle lead to an error of more than a PSF on the position of the planet (see Equation~\ref{eq:delta_mass}).  This could be explained by the fact that the algorithm somehow compensates for the wrong mass by altering some of the orbital parameters. Figure~\ref{Fig:orbit_param_delta_mass}, which gives the retrieved orbital parameters as a function of $\delta{}M$, indicates that both $a$ and $e$ are strongly correlated with $\delta{}M$, and thus that varying these parameters can indeed help to compensate for the error on the stellar mass.

Although this has not been studied in depth in this work, Figure~\ref{Fig:dist_delta_mass} indicates that K-Stacker can potentially be used to retrieve the mass of the central star together with the other parameters. The maximum S/N value of $\snks = 6.34$ is obtained at $\delta{}M = 0$ and, in the region of small mass errors ($\delta{}M/M<10\%$), the $\snks$ function shows a clear drop of up to 0.2 to 0.4 when departing from $\delta{}M = 0$, with no other apparent local maximum. The situation is more complicated for higher initial error on the mass, with a distinct secondary maximum of $\snks$ around $\delta{}M = 0.2\,M_\odot{}$ visible in Figure~\ref{Fig:dist_delta_mass}. The reason for the existence of such a secondary maximum is not clear, and the grid sampling could play an important role here. Furthermore, it remains to be determined whether this apparent local maximum exists in the full parameter space, or if it is an effect of the projection of the $\snks$ function against $\delta{}M$ only. 
A gradient descent taking into account all the orbital parameters and the stellar mass simultaneously could potentially avoid this apparent secondary maximum.
But this remains to be demonstrated, and a significant re-work of the algorithm is necessary to constrain the stellar mass on targets for which it is largely uncertain. 
This is out of scope of the current paper but could be considered, if necessary, for example in the case of low mass stars where the mass is not always well known.

\subsubsection{Tolerance on the true north error}

The True North (TN) gives the absolute rotational orientation of observations. An error on the true north is responsible for a rotational misalignment between the different images used by K-Stacker, resulting in an apparent deviation of the planet motion from Keplerian orbits. For reference, the typical true north error in SPHERE, estimated using a well defined observing strategy of a given astrometric field \citepads{2016SPIE.9908E..34M}, does not exceed 0.1~deg. For NaCo, similar values of $0.1-0.2$\,deg were obtained over more than a decade \citep{2012A&A...542A..41C, 2015A&A...573A.127C}.  
To check the consequences of a TN error on the K-Stacker algorithm, we simulated several observations taking into account the real distribution of the TN errors measured on SPHERE (see Tab. \ref{Tab:TN_error_distribution}). 
  The maximum TN error is $0.08\, \mathrm{deg}$, with a standard deviation of $0.026 \, \mathrm{deg}$, which corresponds to $0.4\, \mathrm{mas}$, or $ \approx 1/100$ of the FWHM of the PSF at the edge of the AO corrected area. At that level, this error should not affect the performance of K-Stacker in any noticeable way. And indeed, we found that K-Stacker had no problem to recover the planets, and gave the same final S/N ratios and orbital parameters, even when multiplying the SPHERE TN errors by a factor of 10.

\subsubsection{Tolerance on the stellar distance error} 

The stellar distances are known from the \textit{Hipparcos} \citepads{2007A&A...474..653V} or \textit{GAIA} missions \citepads{2018A&A...616A...9L}. The typical error on the parallax given by \textit{GAIA} DR2 for bright sources (mag < 14) is $\delta{}\varpi < 0.1\,\mathrm{mas}$ \citepads{2018A&A...616A...9L}. For a star at 10 pc ($\varpi = 100~\mathrm{mas}$), this translates to a relative error on the distance of $\delta{}d/d = \delta{}\varpi/\varpi = 0.1\%$. 

To determine the position of the possible planet in each image, K-Stacker first calculates the position of the planet around its central star, and then projects this position on the detector. The projected position is directly proportional to $d^{-1}$. Given that the AO corrected field on SPHERE is typically $1 ''$, the projection error induced by a 0.1\% error on the distance of the star can be at most 1 mas (i.e., much smaller than the instrumental PSF). Consequently, this error is negligible for K-Stacker.

 \section{Applying K-Stacker to known exoplanets} \label{search for shine companions}

\begin{table*}[h]
\begin{center}  
 \caption{Search space for the two K-Stacker runs on two targets of the SHINE survey.}
  \begin{tabular}{c |l l| l l}
    \hline
    \hline
    Parameter & \multicolumn{2}{c|}{$\beta$ Pictoris} & \multicolumn{2}{c}{HD 95086} \\
    & Range & Distribution & Range & Distribution \\
    \hline
    $M_{\text{star}}$ & $1.75$ M$_\odot{}$ & fixed value & $1.6$ M$_\odot{}$ & fixed value \\
    $d_{\text{star}}$ &  19.75 pc & fixed value &  83.8 pc & fixed value  \\
    $a$ & [2.5 au,  13 au] & uniform & [40 au, 63 au] & uniform \\
    $e$ & [0, 0.8]  & uniform & [0, 0.8]  & uniform \\
    $t_0$ & [ 0 yr, 37 yr] & uniform & [ 0 yr, 386 yr] & uniform \\
    $\Omega$ & [-180 deg, 180 deg] & uniform & [-180 deg, 180 deg] & uniform \\
    $i$ & [0, 180 deg] & uniform & [0, 180 deg] & uniform \\
    $\theta_0$ & [-180 deg, 180 deg] & uniform & [-180 deg, 180 deg] & uniform \\
    \hline
  \end{tabular}
  \label{tab:search}
\end{center}  
\end{table*}

\begin{figure}[h]
\begin{center}
   \includegraphics[width=7cm]{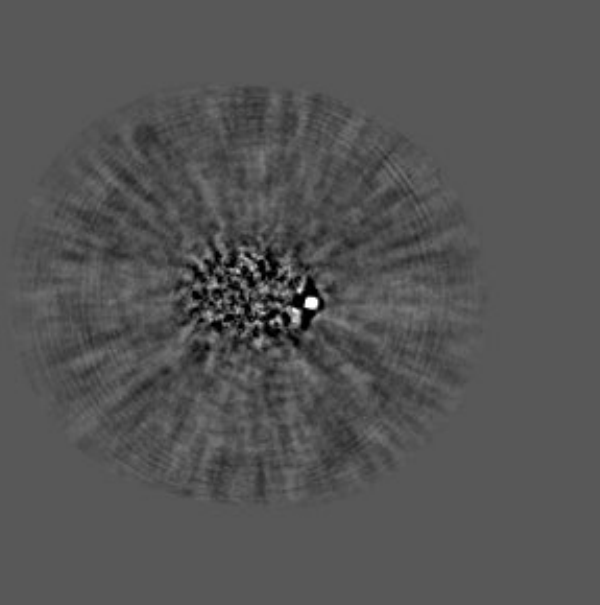}
   \caption{Best recombined image resulting from the K-Stacker run on $\beta$ Pictoris: at each epoch, the images are rotated and shifted to put the planet on its periastron position found by K-Stacker, and the frames are co-added. The planet b is detected at a $\snks$ level of 24.5.}
     \label{Fig:K-Stackerbetapicb}
     \end{center}   
\end{figure}

\begin{figure*}[h]
\includegraphics[width=18cm]{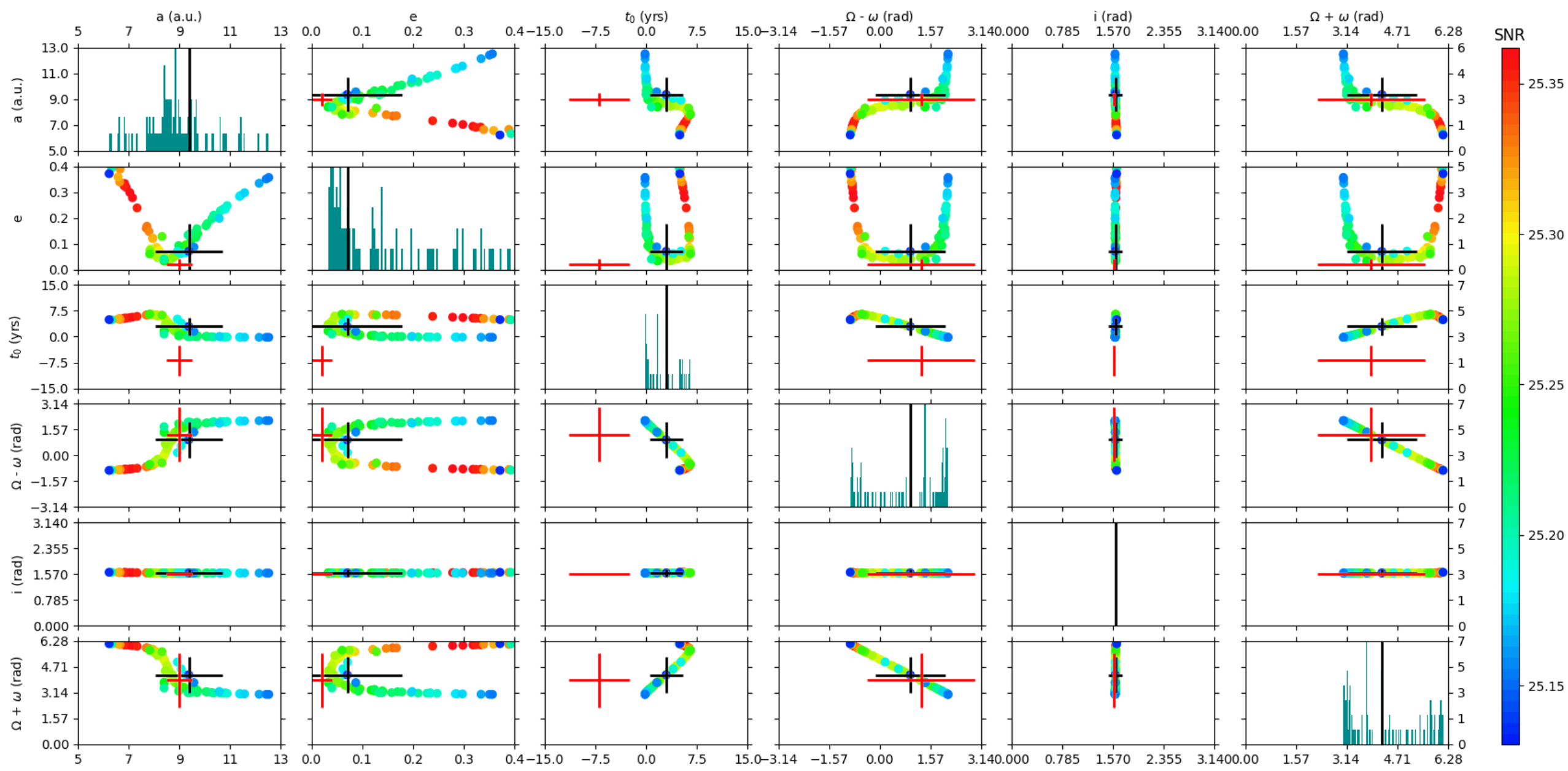}
\caption{Histograms and 2D diagrams of the $\beta$ Pictoris b orbital parameters found by K-Stacker. At left, top and bottom: scale of the orbital parameters. At right: scale of the histograms. 
89 points in each 2D diagram corresponding to the 89 orbits with the highest $\snks$ found by K-Stacker. The color of each point gives the K-Stacker signal to noise indicated at right. The dark cross indicates the mean value of the orbital parameters with their error bars.
The red cross shows the higher probability density found by an MCMC technic in \citepads{2019A&A...621L...8L} converted in the K-Stacker referential. The origin of the t$_0$ K-Stacker date is the $01/12/2014$.} 
\label{Fig_orbit_beta_pic_b}
\end{figure*}

\begin{figure}[h]
\begin{center}
   \includegraphics[width=7cm]{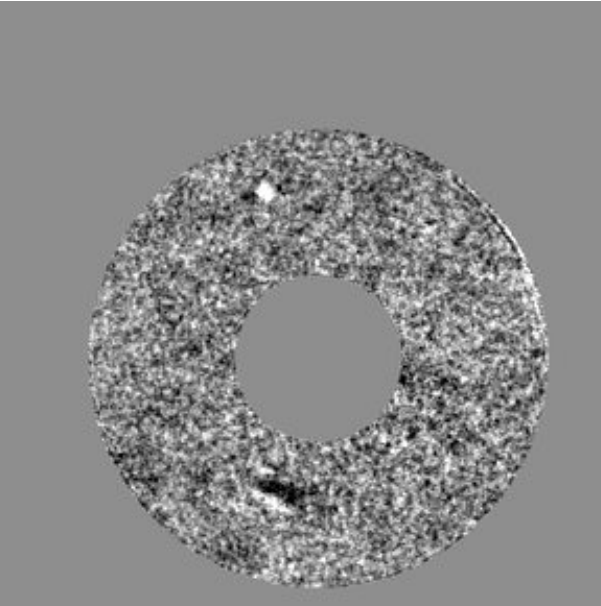}
   \caption{Best recombined image resulting from the K-Stacker run on HD 95086: at each epoch, the images are rotated and shifted to put the planet on its periastron position found by K-Stacker, and the frames are co-added. The planet b is detected at a $\snks$ level of 9.97.}
     \label{Fig:K-StackerHD95086b}
     \end{center}   
\end{figure} 

\begin{figure*}[h]
\includegraphics[width=18cm]{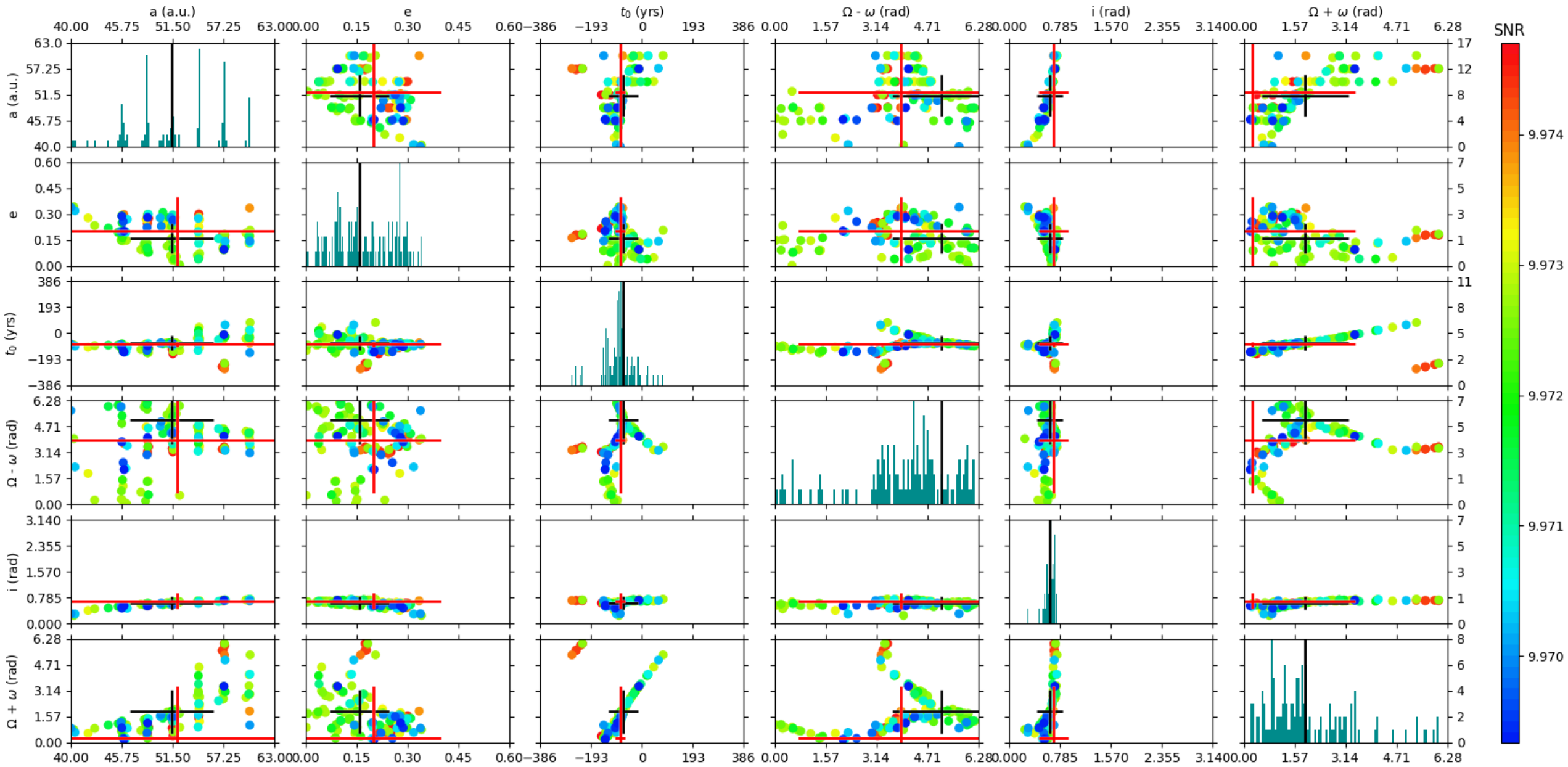}
\caption{Histograms and 2D diagrams of the HD 95086 b orbital parameters found by K-Stacker. At left, top and bottom: scale of the orbital parameters. At right: scale  of the histograms. 
129 points in each 2D diagram corresponding to the 129 orbits of the higher $\snks$  found by K-Stacker. The color of each point gives the K-Stacker signal to noise indicated at right. The dark cross indicates the mean value of the orbital parameters with their error bars.
The red cross shows the higher probability density found by an MCMC technic in \citetads{2018A&A...617A..76C} converted in the K-Stacker referential. The origin of the t$_0$ K-Stacker date is the $05/05/2015$.} 
\label{Fig_orbit_hd95086_b}
\end{figure*} 

\begin{table}
\begin{center}  
 \caption{Mean orbital solutions found by K-Stacker on the two targets of the SHINE survey presented in Section~\ref{search for shine companions} of this paper. The parameter $t_0$ gives the time at periastron, counted from an arbitrary reference date set at 01/12/2014 for $\beta$ Pictoris and 05/05/2015 for HD 95086.}
  \begin{tabular}{c c c c}
    \hline
    \hline
    Parameter & Unit & $\beta$ Pictoris b & HD 95086 b \\
    \hline
    $a$ &  au & $9.37\pm1.68$ & $51.45\pm4.66$ \\
    $e$ & - & $0.07\pm0.12$ & $0.16\pm0.09$ \\
    $t_0$ & yr & $2.95\pm4.78$ & $-70.98\pm55.01$ \\
    $\Omega + \theta_0$ & rad & $4.23\pm1.89$ & $1.89\pm1.33$ \\
    $i$&  rad & $1.60\pm0.11$ & $0.63\pm0.2$ \\
    $\Omega - \theta_0$& rad & $0.95\pm1.89$ & $5.14\pm1.5$ \\
    \hline
  \end{tabular}
  \label{tab:parametersfound}
\end{center}  
\end{table}

In this Section, we present the first results obtained with K-Stacker on two real planets: $\beta$ Pic\,b and HD\,95086\,b. Each of these emblematic objects has been observed multiple times with SPHERE during the SHINE survey, and together they provide a good test of K-Stacker in different conditions. $\beta$ Pic b, is on an edge-on orbit, with a significant orbital motion. HD\,95086\,b is rather on pole-on configuration and moving by only a few PSFs over the 5 years of IRDIS monitoring.

\begin{table*}[h]
\begin{center}  
 \caption{Search space in the inner part of the two stars $\beta\,$ Pictoris and HD 95086.}
  \begin{tabular}{c |l l| l l}
    \hline
    \hline
    Parameter & \multicolumn{2}{c|}{$\beta$ Pictoris} & \multicolumn{2}{c}{HD 95086} \\
    & Range & Distribution & Range & Distribution \\
    \hline
    $M_{\text{star}}$ & $1.75$ M$_\odot{}$ & fixed value & $1.6$ M$_\odot{}$ & fixed value \\
    $d_{\text{star}}$ &  19.75 pc & fixed value &  83.8 pc & fixed value  \\
    $a$ & [2.4 au, 3.5 au] & uniform & [10 au, 22 au] & uniform \\
    $e$ & [0, 0.5]  & uniform & [0, 0.5]  & uniform \\
    $t_0$ & [ 0 yr, 6 yr] & uniform & [ 0 yr, 90 yr] & uniform \\
    $\Omega$ & [-180 deg, 180 deg] & uniform & [-180 deg, 180 deg] & uniform \\
    $i$ & [0, 180 deg] & uniform & [0, 180 deg] & uniform \\
    $\theta_0$ & [-180 deg, 180 deg] & uniform & [-180 deg, 180 deg] & uniform \\
    \hline
  \end{tabular}
 
  \label{tab:searchc}
\end{center}  
\end{table*}

\begin{figure*}[!h]
\begin{center}
   \includegraphics[width=0.8\linewidth]{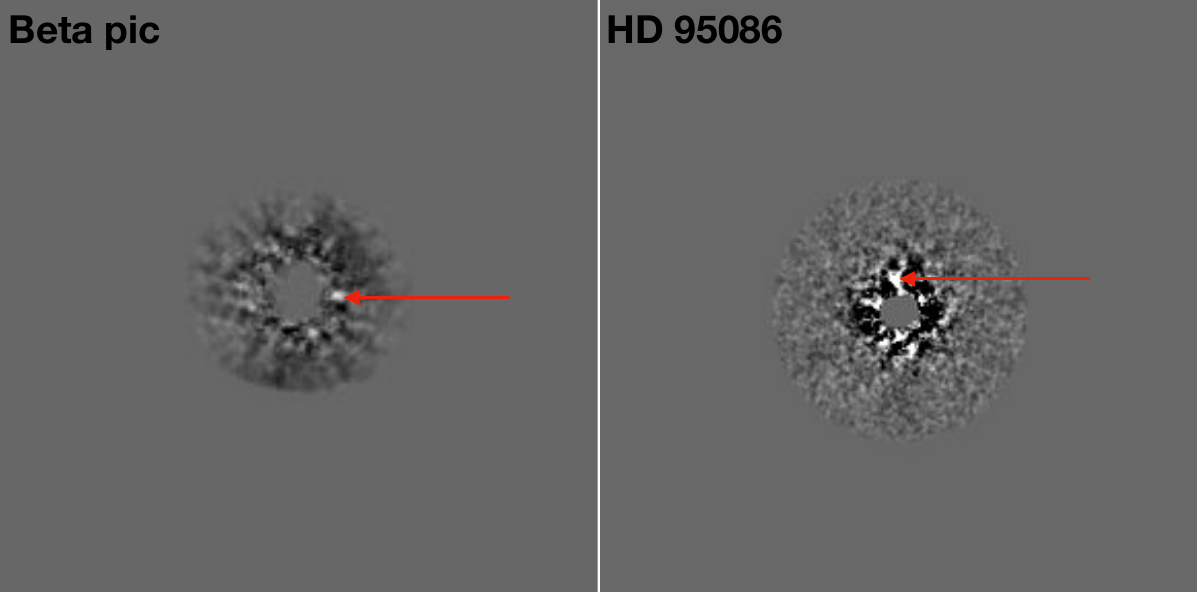}
   \caption{Recombined images obtained with K-Stacker, when searching for additional companions around $\beta$ Pictoris and  HD~95086, two systems observed in the SHINE survey. 
   At each epoch, the images are rotated and shifted to put the detection on its periastron position found by K-Stacker, and the frames are co-added.
   In each case, a bright spot can be seen near the coronagraphic mask (red arrow), which corresponds to the best solution found by the algorithm. In the case of $\beta$ Pic, the corresponding $\snks$ is 4.9. For HD~95086, the $\snks$ is 4.4.}
     \label{shine_c}
     \end{center}
\end{figure*} 

\subsection{$\beta$ Pic b} \label{betapicb}

Eleven IFS observations of $\beta$ Pictoris, spread over more than three years between 2015 and 2018, were available in the SHINE survey (Table \ref{tab:observations}). We reduced these data with a PCA ASDI algorithm \citepads{2015A&A...576A.121M}.
Although in this case, with a mean S/N ratio of 7.42, the planet is clearly detected in each individual image, K-Stacker was set up to look blindly for planets in the range of orbital parameters given in Table~\ref{tab:search}. 

The planet $\beta$ Pic b was detected at a total K-Stacker recombined $\snks = 24.5$ (see Fig~\ref{Fig:K-Stackerbetapicb}), a gain of a factor 3.3 compared to the individual PCA ASDI reduced observations. For a set of 11 observations, this gain is optimal.

Figure \ref{Fig_orbit_beta_pic_b} gives the distribution of the 89 best orbits found by K-Stacker in the parameter space. The black crosses on the different sub-plots give the position of the mean value of these 89 orbits, with the associated $1\,\sigma$ spread (see also Table \ref{tab:parametersfound}). For comparison, the red cross gives the best estimates and $1\,\sigma$ uncertainties from \citetads{2019A&A...621L...8L}, converted to the reference system used in K-Stacker.

Since K-Stacker does not implement a proper MCMC exploration of the parameter space, the statistical meaning of the corner plots presented in Fig. \ref{Fig_orbit_beta_pic_b} is not straightforward. But these pseudo-corner plots share some interesting similarities with the results of a more classical approach to fitting, as presented in \citetads{2019A&A...621L...8L}: a clear V shaped correlation between the semi-major axis $a$ and the eccentricity $e$, related to a degeneracy on the position of periastron/apoastron; a well constrained edge-on inclination; and an eccentricity distribution which peaks at $e<0.1$. Overall, the orbital solution resulting from the K-Stacker run is in good agreement with the recent results of \citetads{2019A&A...621L...8L}. The only significant difference is on $t_0$ found by \citetads{2019A&A...621L...8L} that is near the apoastron of the K-Stacker mean solution (see red and dark crosses for $t_0$ in Fig. \ref{Fig_orbit_beta_pic_b}). This ambiguity in the periastron/apoastron is reinforced by the small eccentricity and the incomplete coverage of the orbit. It will be solved with further monitoring.

\subsection{HD\,95086\,b} \label{HD95b}

For HD~95086, a total of eight observations are available in the SHINE survey between May 2015 and May 2019, and all obtained in good conditions in the $H$ part of the spectrum with IFS (see Table \ref{tab:observations}). The data were reduced using a PCA ASDI algorithm. 


The algorithm was set to search for planets in the range of parameters given in Table~\ref{tab:search}. The planet HD~95086~b is detected (Fig. \ref{Fig:K-StackerHD95086b}) at a recombined $\snks = 9.97$. The mean S/N in the individual PCA ASDI reduced observations was 3.67. Thus, the signal to noise was improved by a factor 2.72 by K-Stacker. For a series of 8 observations, this is again very close to the optimal case. The mean orbital solution found by K-Stacker is presented in Table~\ref{tab:parametersfound}, and the associated pseudo-corner plot can be found in Fig~\ref{Fig_orbit_hd95086_b}. 
Although 1.4 $\%$ of the orbital period of HD 95086 b has been covered by the observations, the orbital parameters are relatively well constrained. 
Again, the pseudo-corner plots of K-Stacker share some similarities with the results of a more classical approach to fitting, as presented in \citetads{2018A&A...617A..76C}.
Within the error bars, the mean values of the orbital parameters found by K-Stacker (see dark and red crosses of Fig. \ref{Fig_orbit_hd95086_b}) are equal to the solutions coming from the MCMC technique described in \citetads{2018A&A...617A..76C}.

 \section{Searching for inner exoplanets} \label{search for new planets}

In this Section, we focus on the search for additional inner planets around HD~95086 and $\beta$ Pictoris using the SHINE reduced data (Table \ref{tab:observations}). Indeed, the presence of one or two additional inner giant planets is suspected considering the double-belt architecture of HD~95086 \citepads{2018A&A...617A..76C} and that a $\sim 9$ M$_\text{Jup}$ planet $\beta$ Pic c has been found using radial velocity, at 2.7 au from the star \citepads{2019NatAs.tmp..421L}.  To search for additional companions in these systems, we proceed using the same K-Stacker algorithm, in which we introduce an initial step of masking the known imaged planet b from the individual images.
We search in the inner area of HD~95086 and $\beta$ Pictoris with the parameters given in the Table \ref{tab:searchc}.

\begin{table}
\begin{center}  
 \caption{Orbital parameters of the solution found by K-Stacker in the inner part of $\beta\,$ Pictoris  }
  \begin{tabular}{c c c }
    \hline
    \hline
    Parameter & Unit &  values  \\
    \hline
    $a$ &  au &  $3.0\, \pm 0.3$ \\
    $e$ & - &  $0.14 \pm 0.1$  \\
    $t_0$ & yr & $-8.3 \, \pm  \, 0.8$  \\
    $\Omega$ & rad & $-2.0 \, \pm 1.5$  \\
    $i$&  rad &   $+1.85 \, \pm 0.86$\\
    $\theta_0$& rad &  $+4.7 \, \pm 1.5$ \\        
    \hline
  \end{tabular}
  \label{Tab_orbit_betapic_c}
\end{center}  
\end{table}

For both HD~95086 and $\beta$ Pictoris, we find a bright secondary feature, close to the coronagraphic mask (see Fig. \ref{shine_c}). 
For HD~95086, the feature is located at $a=16.7$ au and has a low $\snks = 4.4$. Its spread shape indicates a probable false positive. 
A chromatic study shows that this bright speckle is a false positive (Desgrange et al. 2020, In preparation). \\

In the case of $\beta$ Pictoris, the compact feature is found at $\snks = 4.9$ (Fig. \ref{shine_c}) with the orbital parameters given at Table \ref{Tab_orbit_betapic_c}.\\

This solution is compatible with the orbital parameters inferred from the radial velocities \citepads{2019NatAs.tmp..421L}: \\

$a=2.6948$ \text{au}, $e=0.243$, $t_0=-7.87206$ \text{yr}, $\theta_0=4.62$ \text{rad} (here $t_0$ and $\theta_0$ were expressed in the K-Stacker referential).\\

The semi-major axis found by K-Stacker is larger by 0.3 au but the PSF of the planet c is partially masked at several epochs by the coronagraphic mask and can therefore disturb the solution of K-Stacker. However, the parameters
 $\Omega =  -2.0 \, \text{rad} \pm 1.5 \, \text{rad}$ and $i=+1.85 \, \text{rad} \pm 0.86 \, \text{rad}$ give an orbit misaligned with the disk that is very difficult to explain from a dynamical point of view. 
In the blind test (Sect. \ref{result_blind_test}), we had two ambiguous cases where the orbits found by K-Stacker passed well by the planet positions only at 3-4 epochs over 10. Thus, even if the orbit found by K-Stacker is not correct, a part of the light of this 'bright' structure (Fig. \ref{shine_c}) could come from $\beta$ Pictoris c.

In all these cases, the probability of a true detection at $\snks \leq 5$ is smaller than $50 \%$.

\section{Discussion on the strategy of future K-Stacker observations} \label{discussion}

A fundamental question for any potential future surveys with K-Stacker will be to select the optimum number of observations per target.
The blind tests realised on simulated IRDIS images \citepads{2018A&A...615A.144N} and on real data (see Sect. \ref{real data blind test and search}) show that an $\snks > 7$ must be reached to claim a true detection with high confidence (i.e. small rate of false alarm).
An $S/N> 7$ is also required to be able to do a precise spectral analysis and confirm a true detection by characterising the physical properties of the planet (atmosphere composition, temperature, surface gravity, clouds detections, etc.); 
see for example the characterisation of the planets around the stars HR8799, $\beta$ Pic, HD95086, 51Eri, HIP65426, PDS70 (\citeads{2016A&A...587A..58B}, \citeads{2017AJ....153..182C}, \citeads{2016ApJ...824..121D}, \citeads{2017A&A...605L...9C}, \citeads{2018A&A...617L...2M}).
But, further work is required to develop a K-Stacker exoplanet statistical analysis tool : using a similar approach than with the multi-purpose exoplanet simulation system (MESS) algorithm \citepads{2012A&A...537A..67B}, we should be able to inject numerous fake planets 
in series of observations in order to compute a probability of detection in function of the planet mass by using model mass-luminosity relationships \citepads{2003A&A...402..701B}. This approach will allow to give an upper mass limit for planets hidden in the data even when K-Stacker does not detect anything.\\

The total exposure time required for K-Stacker will also depend of the number of observations already done. For example, to confirm the detection of $\beta$ Pictoris c that has been found at  $\snks = 4.8$ in eleven exposures, 
a minimum of $11*[(7/4.8)^2 - 1] \approx 12$ new observations are needed (perhaps a little less by taking care 
to observe $\beta$ Pictoris c when it is fully out of the coronagraphic mask using radial velocity constraints) to reach $\snks \approx 7$. \\

Fig. \ref{Fig:snr_ks_tot} and Table \ref{Tab_orbits} shows that K-Stacker is able to recombine in a close to optimal way, series of observations with very different orbital parameters. 
Thus, in principle, the total S/N of a K-Stacker run should only depend on the total exposure time, and not on the number of individual exposures in which it is divided. 
In these conditions, better constraints on the orbital parameters could be obtained at fixed total exposure time by taking as many exposures over a period as long as possible. 
But, to reach higher contrast (i.e. per individual exposure and in the final K-Stacker recombined image), a minimum exposure time by observation is required to have enough field of view rotation for a good ADI substraction (i.e. in each observation, a parallactic rotation of at least one PSF FWHM  at the position of the planet
 must be reached to subtract efficiently the instrumental speckles with ADI).
The trade-off comes from the difficulty in setting up the observation strategy, the telescope overheads, and the star declination (to adjust the minimum exposure of individual observations).\\

For future imaging surveys with K-Stacker on SPHERE+ \citepads{2020arXiv200305714B} and/or the ELTs instruments, a software to optimize observing schedules will have to be developed. 
Inspired from the algorithm used in the SHINE survey \citepads{2016SPIE.9910E..33L}, the goal of this K-Stacker scheduler will be to compute the minimum exposure time
required at each epoch for each star to have an efficient ADI subtraction, but at the same time to split the observations over at least $10-20 \%$ of the orbital period of the searched planets to get accurate orbital parameters.

\section{Conclusions} \label{conclusion}

For the first time, we tested the K-Stacker algorithm on real IRDIS and IFS observations, reduced with the PCA ADI and ASDI algorithms, in a similar fashion as in the SHINE survey.

From a blind experiment, in which planets are injected on random orbits in the images before the PCA ADI reduction, and recovered by the K-Stacker algorithm, we concluded that the detection statistics are similar to what was previously obtained with simulated non-ADI reduced images: the success rate is close to 100\% when searching for companions for which the recombined $\snks$ is $\simeq{9}$, and it drops significantly at $\snks\simeq{5}$.

Using data on two targets repeatedly observed during the SHINE survey ($\beta$ Pic, observed 11 times; HD~95086, observed 8 times), we have shown that the K-Stacker algorithm  is good at recovering the known companions $\beta$~Pic~b and HD~95086~b. K-Stacker also provides orbital parameter estimates in agreement with current literature values. The gain provided by K-Stacker recombination is very close to the square-root of the number of observations combined (i.e. close to optimal).

We also searched for additional sub-stellar companions around these two stars. We found two bright features, corresponding to two possible planets orbiting within the orbit of the known companions. However, these two features are close to the coronagraphic mask, and we will show in a dedicated analysis that the one found around HD~95086 is a recombined "super-speckle". The c candidate detected by K-Stacker around $\beta$ Pictoris without using prior information from the radial velocity detection of $\beta$ Pictoris  c is on a trajectory compatible with the orbital parameters found by \citetads{2019NatAs.tmp..421L}. But, the K-Stacker orbit is misaligned with the disk and the $\snks \approx 5$ is not high enough to claim that it is a true detection. Despite the relatively large error bars on the euler angles ($\Omega$, $i$, $\theta_0$) the $1-\sigma$ agreement between the putative K-stacker detection and the radial velocity solution is encouraging and more observations are required to constrain the orbital parameters and to determine if at least a part of the light of this detection comes from $\beta$ Pictoris c.

\begin{acknowledgements}

SPHERE is an instrument designed and built by a consortium
consisting of IPAG (Grenoble, France), MPIA (Heidelberg, Germany), LAM
(Marseille, France), LESIA (Paris, France), Laboratoire Lagrange
(Nice, France), INAF–Osservatorio di Padova (Italy), Observatoire de
Genève (Switzerland), ETH Zurich (Switzerland), NOVA (Netherlands),
ONERA (France) and ASTRON (Netherlands) in collaboration with
ESO. SPHERE was funded by ESO, with additional contributions from CNRS
(France), MPIA (Germany), INAF (Italy), FINES (Switzerland) and NOVA
(Netherlands).  SPHERE also received funding from the European
Commission Sixth and Seventh Framework Programmes as part of the
Optical Infrared Coordination Network for Astronomy (OPTICON) under
grant number RII3-Ct-2004-001566 for FP6 (2004–2008), grant number
226604 for FP7 (2009–2012) and grant number 312430 for FP7
(2013–2016). We also acknowledge financial support from the Programme National de
Planétologie (PNP) and the Programme National de Physique Stellaire
(PNPS) of CNRS-INSU in France. This work has also been supported by a grant from
the French Labex OSUG@2020 (Investissements d’avenir – ANR10 LABX56).
The project is supported by CNRS, by the Agence Nationale de la
Recherche (ANR-14-CE33-0018). It has also been carried out within the frame of the National Centre for Competence in 
Research PlanetS supported by the Swiss National Science Foundation (SNSF). MRM, HMS, and SD are pleased 
to acknowledge this financial support of the SNSF. Finally, this work has made use of the the SPHERE
Data Centre, jointly operated by OSUG/IPAG (Grenoble),
PYTHEAS/LAM/CESAM (Marseille), OCA/Lagrange (Nice), Observatoire de
Paris/LESIA (Paris), and Observatoire de Lyon, also supported by a grant from Labex 
OSUG@2020 (Investissements d’avenir – ANR10 LABX56). We thank P. Delorme and E. Lagadec (SPHERE Data
Centre) for their efficient help during the data reduction
process. This research has made use of computing facilities operated by CeSAM data center at LAM, Marseille, France.

\end{acknowledgements}

%
%
\bibliographystyle{aa} 
\bibliography{lecoroller_biblio}   

\begin{appendix}
\onecolumn
\section{SHINE blind test}

\begin{table}[h]
\begin{center}
	\begin{tabular}{|c|c|c|c|c|c|c|c|c|c|}
	\hline 
	 & $a$  (au) & $e$  & $t_0$  (Year) & $\Omega$  (rad) & $i$ (rad)  & $w_0$ (rad) & $\Delta$ (pixels) & $S/N$ & $\%T $ \\
	\hline 
	\hline
	Injected & 4.17  & 0.25  &  -0.84 & 1.9 & 2.89 &  0.08 & & 11  & 0.38 \\  

	Found & 4.54 $\pm$  0.45 & 0.19 $\pm$  0.03 &  -0.62 $\pm$   0.37 & 2.35 $\pm$  0.66 & 2.73 $\pm$ 0.11& 0.93 $\pm$ 0.66 & 0.41 & 10.5 &\\  
	\hline
	
	Injected & 6.24 & 0.41 & -0.92 & 3.33  & 2.25 & 2.19 & & 9.9 & 0.21\\ 
 
	 Found & 4.91  $\pm$ 0.5 &  0.37 $\pm$ 0.03 & -1.4 $\pm$ 0.22  & 3.47 $ \pm$ 0.2 & 2.35 $\pm$ 0.11 & 1.76 $\pm$ 0.2 & 0.40 &  9.4 &\\ 
	\hline 
	
	 Injected &  7.37 &  0.19 & -7.14 & 6.89 & 0.58 & 3.99 & & 8 &  0.15\\ 
	
	  Found & 7.02 $\pm$  0.26 & 0.34 $\pm$ 0.03 & -7.70 $\pm$ 0.83 & 6.15  $\pm$ 0.33 &  0.53 $\pm$ 0.11 & 4.43 $\pm$ 0.33 & 0.65 &  8.28&\\
	\hline
	
	Injected & 6.51 & 0.07&19.63&0.06&2.69&6.81&& 8.2 & 0.19\\ 
	
	Found  & 6.18 $\pm$  0.39 & 0.01 $\pm$ 0.06 & 13.48 $\pm$ 3.13 & 0.39 $\pm$ 1.43 & 2.73 $\pm$ 0.17 & 5.11 $\pm$ 1.43 & 0.27 & 8.3	 & \\ 
	\hline
		
	Injected & 5.72 &  0.1 &  2.76 &  2.80 & 2.83 & 4.62 & & 6.47 & 0.23 \\ 
	
	Found  & 6.3 $\pm$ 0.48 & 0.17 $\pm$ 0.04 &  2.04 $\pm$ 0.49 & 2.62 $\pm$ 1.01 & 2.74 $\pm$ 0.16 & 4.05 $\pm$ 1.01 & 0.45 & 6.3 &  \\ %
	\hline		
	\end{tabular}
\end{center}
\caption{Orbital parameters of the injected and  found planets (true positives) during the blind test described at Sect. \ref{real data blind test and search}. 
$\Delta$ (pixels) is the mean distance between the injected and found positions of the planet in the images. 
$S/N$ is the signal to noise computed using the orbits of injection $\sntot$ and the mean orbit found $\snks$.
$\%T $ is the percentage of the covered orbit. }
\label{Tab_orbits}
\end{table}

\begin{table}[h]
\begin{center}
\begin{tabular}{|c|c|}
\hline
 Date (Yr) & True North (deg) \\
\hline
 0. & -1.813 \\
\hline
 0.1 & -1.795 \\
\hline
 0.2 & -1.758 \\
\hline
 0.3 & -1.749 \\
\hline
 1.1 & -1.761 \\
\hline
 1.2 & -1.759 \\
\hline
 2.1 & -1.739 \\
\hline
 2.2 & -1.773 \\
\hline
 3.1 & -1.8024 \\
\hline
 3.2 & -1.804 \\
\hline
\end{tabular}
\end{center}
\caption{Distribution of the True North errors measured with SPHERE 
 on the 47 Tucanae field \citepads{2016SPIE.9908E..34M}.}
 \label{Tab:TN_error_distribution}
\end{table}

\section{SHINE IFS observations}
\begin{table}[h]
\begin{center}  
  \begin{tabular}{|c|c|c|c|c|c|c|c|c|c|}
    \hline
    \hline
    Star name & Obs. date & JD & NDIT$\times$DIT & Rot & Seeing & Prism & Coro & algorithm & modes \\
    \hline
$\beta$~Pic & 2015 Feb 05 & 57058.02 & 316$\times$8 & 88.36 & 0.89 & Y - J & APO1-ALC2 & ASDI-PCA & 50 \\
$\beta$~Pic & 2015 Sep 30 & 57296.33 & 318$\times$8 & 36.44 & 1.37 & Y - J & APO1-ALC2 & ASDI-PCA & 50\\ 
$\beta$~Pic & 2015 Nov 30 & 57356.23 & 880$\times$4 & 39.72 & 1.85 & Y - J & APO1-ALC2 & ASDI-PCA & 50\\ 
$\beta$~Pic & 2015 Dec 26 & 57382.15 & 223$\times$16 & 37.40 & 0.99 & Y - J & APO1-ALC2 & ASDI-PCA & 50\\ 
$\beta$~Pic & 2016 Jan 20 & 57407.10 & 160$\times$16 & 29.54 & 1.07 & Y - J & APO1-ALC2 & ASDI-PCA & 50\\ 
$\beta$~Pic & 2016 Apr 15 & 57493.97 & 168$\times$16 & 20.27 & 0.75 & Y - J & APO1-ALC2 & ASDI-PCA & 50\\ 
$\beta$~Pic & 2016 Sep 16 & 57647.36 & 348$\times$16 & 38.77 & 0.78 & Y - J & APO1-ALC2 & ASDI-PCA & 50\\ 
$\beta$~Pic & 2016 Oct 14 & 57675.32 & 380$\times$16 & 53.03 & 0.73 & Y - J & APO1-ALC2 & ASDI-PCA & 50\\ 
$\beta$~Pic & 2016 Nov 18 & 57710.25 & 564$\times$4 & 44.20 & 0.78 & Y - J & APO1-ALC2 & ASDI-PCA & 50\\ 
$\beta$~Pic & 2018 Sep 17 & 58378.35 & 376$\times$8 & 36.42 & 0.87 & Y - J & APO1-ALC2 & ASDI-PCA & 50\\ 
$\beta$~Pic & 2018 Oct 18 & 58409.31 & 432$\times$8 & 54.40 & 0.78 & Y - J & APO1-ALC2 & ASDI-PCA & 50\\
HD 95086 & 2015 May 05 & 57147.04 & 47$\times$64 & 17.42 & 0.72 & Y - H & APO1-ALC2 &ASDI-PCA & 50\\
HD 95086 & 2016 May 30 & 57538.97 & 61$\times$64 & 25.63 & 0.48 & Y - H & APO1-ALC3 &ASDI-PCA & 50\\ 
HD 95086 & 2017 May 09 & 57882.96 & 99$\times$64 & 36.55 & 0.94 & Y - H & APO1-ALC3 &ASDI-PCA & 50\\  
HD 95086 & 2018 Jan 06 & 58124.30 & 70$\times$64 & 41.05 & 0.30 & Y - H & APO1-ALC3 &ASDI-PCA & 50\\ 
HD 95086 & 2018 Feb 24 & 58173.19 & 64$\times$96 & 33.59 & 0.32 & Y - H & APO1-ALC3 &ASDI-PCA & 50\\ 
HD 95086 & 2018 Mar 28 & 58205.12 & 64$\times$96 & 33.45 & 0.57 & Y - H & APO1-ALC2 &ASDI-PCA & 50 \\ 
HD 95086 & 2019 Apr 13 & 58586.07 & 63$\times$96 & 33.84 & 1.05 & Y - H & APO1-ALC2 &ASDI-PCA & 50\\ 
HD 95086 & 2019 May 17 & 58620.99 & 64$\times$96 & 33.29 & 0.84 & Y - H & APO1-ALC2 &ASDI-PCA & 50\\ 	
    \hline
  \end{tabular}
  \caption{Observations used in this paper.}
  \label{tab:observations}
\end{center}  
\end{table}

\end{appendix}

\end{document}